\documentclass[12pt]{article}
\usepackage{jheppub}
\usepackage{amsmath,amssymb}
\usepackage{slashed,verbatim,graphicx}
\usepackage{graphicx}
\usepackage{ytableau}

\usepackage{color}

\newcommand\vev[1]{\langle #1\rangle}
\newcommand\ket[1]{| #1\rangle}

\newcommand{\vac}{\ket 0}

\newcommand\BR{\mathbb{R}}

\def\Tr{\textrm{Tr}}

\begin{document}

\title{Submatrix deconfinement and small black holes in AdS}

\author{David Berenstein}\emailAdd{dberens@physics.ucsb.edu }

\affiliation{ Department of Physics, University of California, Broida Hall, Bldg 572, Santa Barbara, CA 93106}

\abstract{Large $N$ gauged multi-matrix quantum mechanical models usually have a first order Hagedorn transition, related to deconfinement.  In this transition the change of the energy and entropy is of order $N^2$ at the critical temperature. 
This paper studies the microcanonical ensemble of the model at intermediate energies $1<<E<<N^2$ in the coexistence region for the first order phase transition. Evidence is provided for a partial deconfinement phase
where  submatrix degrees of freedom for a $U(M)$ subgroup of $U(N)$, with  $M<<N$ have an excitation energy of order $M^2$ and are effectively phase separated from the other degrees of freedom. These results also provide a simple example of the
 Susskind-Horowitz-Polchinski correspondence principle where a transition from a long string to a black hole is smooth. Implications for the dual configurations of small black holes in $AdS$ are discussed. 
}

\maketitle

\section{Introduction}

In statistical mechanics, to define a phase transition usually requires first taking the thermodynamic limit of infinite volume. In a quantum theory at finite volume one would usually expect that the spectrum of states is discrete and that phase transitions are smoothed out into a rapid crossover. However, when we consider the large $N$ limit of field theories, phase transitions can show up, even at finite volume.  This is usually exemplified by the Gross-Witten phase transition \cite{Gross:1980he}.

A more standard phase transition that can appear at infinite $N$ is the confinement/deconfinement phase transition of gauged theories in $0+1$ dimensions. 
This transition can arise from doing large $N$ quantum mechanics theories on a compact space at finite temperature.  This phase transition can many times be understood in terms of 
a Hagedorn density of states \cite{Hagedorn:1968zz}, where the density of states grows exponentially in the energy, and it is usually first order.

At low temperatures, the gauge degrees of freedom are confined. Here we assume the gap from the ground state to the first excited state is of order one.
The entropy and  energy are then of order one at low temperatures. At high temperatures, all the degrees of freedom participate, and the 
 energy and entropy are of order $N^2$. A special example of this phenomenon  can be found in holography, where we take ${\cal N}=4 $ SYM theory on $S^3\times \BR$.
 At low temperatures, the theory can be dimensionally reduced on the sphere and one can think of it as a gauged matrix quantum mechanics with a gap. This theory confines and in the weak coupling limit has a Hagedorn density of states, where one can analyze the phase transition perturbatively in the coupling constant \cite{Aharony:2003sx}.
 At strong coupling, this is better described by the AdS/CFT correspondence. The confinement/deconfinement phase transition \cite{Witten:1998zw} can be seen in the dual gravity theory in terms of the Hawking-Page phase transition for black holes in AdS \cite{Hawking:1982dh}. This is a first order phase transition with critical temperature of order one.
 
 Both the entropy and the energy of the black hole are of order $N^2$ (this is normalized by the Newton constant in $AdS_5$ which is of order $N^{-2}$). This matches the expected field theory behavior.  This transition is usually described in terms of a Euclidean path integral where one fixes the temperature and which corresponds to a canonical ensemble in statistical mechanics.
 
 A rather important point for first order phase transitions is that they allow for a coexistence phase. The coexistence phase is usually
 described by the Maxwell construction in the phase diagram. There is usually a geometric phase separation in the coexistence phase. For example, in the liquid-gas phase transition,
 the liquid and the gas are separated spatially. A liquid drop forms a sphere whose shape is controlled by the surface tension of the liquid-gas interface. The simplest ensemble that describes the coexistence phase is the microcanonical ensemble:  one fixes the energy (and the volume) rather than the temperature. 
 
 This behavior of volumes of gases in a coexistence phase prompts the following question: is there a coexistence phase in the Hagedorn transition for matrix quantum mechanics? 
 And if the answer is yes, how can one effectively describe the nature of the phase separation?

 This question is asked in $0+1$ dimensional quantum field theory. Here it is important to note that the phase separation can not occur spatially in a 0-dimensional space, so it must be associated to some information about how the matrix degrees of freedom are related to each other.
 
 In some holographic models, it has been argued that the small black hole phase in $AdS$ is dual to a deconfined phase on a  submatrix set of degrees of freedom
 \cite{Asplund:2008xd, Hanada:2016pwv}. This has not been proven, and alternative options for the dynamical behavior of such theories have been explored by Yaffe \cite{Yaffe:2017axl}.
 
 Indeed, the ideas that led to the sub-matrix proposal by Asplund and the author  \cite{Asplund:2008xd}, depended on a model for the ground state physics at strong coupling that was based on commuting matrices.  Such a model has a good notion of 'density of eigenvalues' at large $N$, which organized itself into the  five sphere $S^5$ \cite{Berenstein:2005aa}. Later considerations showed that such a model probably misses a lot of the physics that can be accessed by other means \cite{Berenstein:2015ooa}. By contrast, the arguments on 
 \cite{Hanada:2016pwv} show that such an interpretation of physics on a submatrix are consistent, but does not show that this is preferred over other possible configurations.
 
 The purpose of this paper is to address this conjecture more seriously from a  first principle set of calculations. We want to address the confinement/deconfinement phase transition at large $N$, before addressing the small black hole phase. Such a Hagedorn transition already exists in the free matrix quantum mechanics limit, bearing in mind that one still needs to impose the singlet 
 constraint. Because of that, if the behavior we are trying to understand is universal, it should be already visible in such an example.
 
 The advantage of such a system is that the states can be readily counted. 
 What makes progress possible to address this conjecture is that there are two ways to count states. One way counts traces. These are interpreted as strings and clearly show the exponential growth of states and the Hagedorn behavior. The physics of the ensemble in this way of counting ends up being dominated by one single long string.
 There is a second way of counting states using combinatorial methods of Young diagrams and group theory. This second way of counting also gives rise to the same counting of states, but the physics can be interpreted differently. Naively, a Young diagram with $n$ rows can be associated to $n$ eigenvalues being separated from $U(N)$. One then shows that the 
 typical state of this ensemble has and energy $E\simeq n_{eff}^2$ and entropy $S\simeq n_{eff}^2$, where $n_{eff}$ is the typical number of rows of the most probable configurations in the ensemble.  This is the same behavior that one would expect for a deconfined phase of $n_{eff}\times n_{eff}$ matrices. 

This observation serves to argue that the answer to the phase separation question is positive: there is a phase separation and the  deconfinement occurs in a submatrix set of degrees of freedom. We will call this property partial deconfinement.

The paper is organized as follows. In section \ref{sec:trace} the free gauge matrix model of two matrices is analyzed and the counting of states is determined in the microcanonical ensemble at large energy. It is shown that the
system has a Hagedorn transition and that a single long trace (string)  dominates the ensemble. In section \ref{sec:eivals} the states in this model are counted by a  different method that utilizes Young diagrams. It is argued that 
in the typical Young diagram the number of rows and columns $n_{eff}$ scale like the square root of the energy, and one can show that the entropy scales like the square $S\simeq n_{eff}^2$. This is used to argue  that the entropy is carried by a submatrix set of degrees of freedom.
This comparison of shape of the Young diagram to the rank of a matrix is a generalization of the intuition that arises from quantum mechanics of matrix models of one matrix. 
 The idea is extended to the study of the duals of small black holes in $AdS$ in section \ref{sec:smallbh}.  To gain control over the dynamics, I study an ensemble that corresponds to ``boosted" black holes in the $S^5$ direction. These carry R-charge. I study the states in a limit where this R-charge dominates and 
 the ensemble can be analyzed either by using spin chain calculations or by computing a gas of massless strings supported on a collection of giant gravitons and dual giant gravitons. It is shown that both of these descriptions seem to give rise to the same physics. This supports the Susskind-Horowitz-Polchinski
intuition that when strings are sufficiently excited they smoothly go to black hole configurations. Finally, in section \ref{sec:conc} I conclude.

\section{Counting states in a simple model: the long trace.}\label{sec:trace}

Consider the simplest quantum mechanical matrix model with two bosonic matrices in the adjoint of $U(N)$: a free model. We are also assuming that  the $U(N)$ dynamics is gauged. It is convenient to express the degrees of freedom 
in terms of raising/lowering operators, so that the excitations of the system are polynomials in $ (a^\dagger)^i_j ,  (b^\dagger)^i_j$, and the vacuum is characterized by $a^i_j\vac= b^i_j\vac=0$. In total we have $2N^2$ degrees of freedom.

For simplicity, the system is chosen so that each of $a,b$ excitations have an energy equal to one $\hbar\omega=1$. In this case the system has an enhanced $SU(2)$ symmetry where $a^\dagger, b^\dagger$ form a doublet. 
This is not essential for this paper. 

The energy of a state is the number of $a,b$ excitations. We will choose to be in a situation where the total energy is fixed and  $1<<E<<N^2$.

As is standard, if we excite the system each $a$ or $b$ excitation carries one upper index and one lower index for $U(N)$. We need to make gauge invariant states, so every upper index needs to be contracted with a  lower index. 
If we follow the index contractions by natural matrix multiplication, we find that in general the gauge invariant states are multi-traces. Each such trace is considered to be a single string.

Let us concentrate on one such trace. We will call the matrix associated to $a^\dagger$, $A$, and the one associated to $b^\dagger$ by $B$. 
We can thus express the traces as traces of a word in the $A,B$ letters.
For example consider
\begin{equation}
\ket{ \ell} \simeq \Tr(ABA\dots )\vac
\end{equation}
where there are exactly $\ell$ letters inside the trace. The energy of this state is exactly $\ell$. 

Since $A,B$ are chosen at random, there are $2^\ell$ possible words inside the trace, but we need to remember that  the traces are cyclic, so if we move the first letter to the last position, we get the 
same state. The order of this group of permutations of the letters is $\ell$, so a better guess for the number of states is
\begin{equation}
{\# \hbox {States}}= 2^\ell/\ell
\end{equation}
There are subleading corrections if identical patterns are repeated. For example, if $\ell= 2k$, we can have a word of $k$ letters that is repeated. Such a configuration would have roughly $2^{\ell/2}$ states,  which are undercounted. After all, they only have $\ell/2$ translates under the cyclic condition, rather than $\ell$. But this is a subleading correction to the number of 
states that is exponentially suppressed.

The entropy we would assign to this collection of (single trace) states is 
\begin{equation}
S= \ell \ln(2) - \ln \ell +O(\exp(-\ell/2)) \simeq \ell \ln(2) - \ln \ell
\end{equation}
If we include multistring states, we must conclude that $S\geq \ell \ln(2) - \ln \ell$ up to exponentially small corrections.

Now, remember that $\ell$ is the energy, so we can view this as a counting of states for a microcanonical ensemble at energy $E= \ell$. The number $\ell$ is also the length of the string (as a word), 
so we can think of strings at large $\ell$ as long strings. 

Using the first law of thermodynamics relation $dE= TdS$ we find that to the single string microcanonical ensemble we can associate a temperature
\begin{equation}
T= \frac{1}{\frac{dS}{dE}}= \frac{1}{\ln(2)-E^{-1}} = \frac{1}{\ln(2)} + \frac{1}{\ln(2)^2 E}+O(1/E^2)
\end{equation}
that asymptotes to $\ln(2)$ at large energies. This is the Hagedorn temperature of the system, $T_H$.  If we had more matrices, let us say $d$ of them,  we would get a similar result 
$S\simeq \ell\ln(d)-\ln \ell$. This would lower the Hagedorn temperature to $T_H\simeq 1/\ln(d)$. This is a simple counting of states that easily shows the Hagedorn behavior. 
A more sophisticated treatment gives the same answer (see \cite{Aharony:2003sx} and references therein).
Most of the results below work equally well with d matrices so long as $d>1$, and we replace $2^\ell$ by $d^\ell$ in the expressions.

Consider now the ensemble of $2$ strings of lengths $\ell_1+\ell_2=\ell$. The number of states in this ensemble would be  given by
\begin{equation}
{\# \hbox {States}}_{\ell_1, \ell_2} = \frac{2^{\ell_1}}{\ell_1}\frac{ 2^{\ell_2}}{\ell_2} = \frac{2^\ell}{\ell_1 \ell_2}
\end{equation}
When we sum over $\ell_1$, we need to be careful, because we can permute the order of the traces. This gives us an extra factor of $1/2$ in the counting. This way we get that  	
\begin{equation}
{\# \hbox {States}}_{2-string} = 2^\ell \frac 12 \sum_{\ell_1+\ell_2=\ell} \frac{1}{\ell_1 \ell_2} =\frac {2^\ell}{2 \ell} \sum_{\ell_1=1}^{\ell-1} \frac{1}{\ell_1}+\frac{1}{\ell-\ell_1}\simeq \frac{2^\ell}{\ell} \ln(\ell)
\end{equation}
There is a subleading correction if both traces have the same size and they are the same. Otherwise the factor of $1/2$  removes the double counting.

This way we get that 
\begin{equation}
{\# \hbox {States}}_{1-string}+{\# \hbox {States}}_{2-string} \simeq \frac{2^\ell}{\ell} \left(1 +\ln(\ell) \right)
\end{equation}
The correction to the entropy from the 2-string states at large $\ell$ is subleading and of the form
\begin{equation}
S_{1+2}\simeq \ell \ln(2) -\ln(\ell) + \ln(1+\ln (\ell))
\end{equation}
so the two string states do not modify the entropy counting by much. A similar phenomenon can be had with the three string states, which will produce a $\ln(\ell)^2$ correction to the total counting of states, to produce
\begin{equation}
{\# \hbox {States}}_{1+2+3-string} \simeq \frac{2^\ell}{\ell} \left(1 +\ln(\ell) +O(\ln(\ell)^2) \right) 
\end{equation}
so that the entropy is
\begin{equation}
S\simeq \ell \ln(2)+ O(\ln(\ell))
\end{equation}
We find this way that the thermodynamic counting of states is dominated by the single long string ensemble, in that we get the correct entropy and the temperature up to subleading logarithmic corrections in the energy. Basically, for most estimates we can work with a single string without worry.

\section{Counting eigenvalues and partial deconfinement }\label{sec:eivals}

There is a second way of counting states at fixed energy $\ell$. This proceeds by realizing that the list of states
\begin{equation}
(a^\dagger)^{i_1}_{j_1} \dots (a^\dagger)^{i_m}_{j_m} \vac
\end{equation}
actually transforms as a representation of $U(N)\times U(N)$, if we are allowed to make rotations in the upper indices independent of each other (this argument has been  used before by the author in \cite{Berenstein:2015ooa}). The states at this stage are not required to satisfy the singlet constraint, which will be imposed later.

 That is, we can assume that the $a^\dagger$ excitations resemble  a
bifundamental representation of $U(N)\times U(N)$. 
 This collection of states can therefore be thought of as a representation of $U(N)\times U(N)$ that can be decomposed into irreducible representations of $U(N)\times U(N)$.
This decomposition proceeds by symmetrizing and antisymmetrizing in the $i_s$ indices according to a Young diagram. The total number of boxes in the Young diagram is equal to $m$.
 Since the
$a^\dagger$ have bosonic statistics, any permutation in the upper indices can be undone by a permutation of the lower indices, so the individual Young diagrams for the $a^\dagger$ in the upper indices is the same as the representation in the lower indices \footnote{If the $a^\dagger$ were fermions there are additional minus signs that flip symmetrizers into antisymmetrizers, so the representation on the lower indices would be associated to the flipped Young diagram \cite{Berenstein:2004hw})}. 
States associated to different Young diagrams are orthogonal to each other automatically, because they correspond to different irreducible representations of $U(N)\times U(N)$, and these are orthogonal to each other. We need to do a similar reasoning with the $b^\dagger$ states. So the total counting of states can be associated to a product of two different Young diagrams with $m$ boxes and $\ell -m$ boxes for each $U(N)$ and summing over all of these choices. 
The original motivation to count states this way arose from studying how to attach strings to giant gravitons \cite{Balasubramanian:2004nb}, which was later formalized via group theory methods in \cite{deMelloKoch:2007rqf}.

The  product tensor obtained this way can again be decomposed into irreducible representations of $U(N)\times U(N)$, which can be analyzed using the Littlewood-Richardson rules. This produces Young diagrams with a total of $\ell$ boxes for the upper indices and lower indices separately. 
Now, we can impose the singlet constraint: the bottom indices and the upper indices can be contracted to a non-zero state if they both produce the same Young diagram. This way one can guarantee that there is a singlet.
It turns out that this procedure also generates a complete counting of states, see \cite{Bhattacharyya:2008rb, Mattioli:2016eyp} and also \cite{Ramgoolam:2016ciq} for a review. To get the result one needs to be careful and take into account that a given representation can have multiplicity.
Adding more matrices requires iterating these ideas, and can be made rather explicit even in quiver setups \cite{Pasukonis:2013ts}.

To illustrate this, consider the product of two tableaux for the $a,b$ oscillators
\begin{eqnarray}
\ytableausetup{boxsize=1.0em}
\ydiagram{2,1}\otimes \ydiagram{2,1}&=&\ydiagram[\bullet]{2+2,1+1}*[]{2,1}+ \ydiagram[\bullet]{2+0,1+1,1,1}*[]{2,1} + \ \ydiagram[\bullet]{2+2,1+0,0+1}*[]{2,1}\\
&+& \ \ydiagram[\bullet]{2+1,1+0,1,0+1}*[]{2,1}+2\ \ydiagram[\bullet]{2+1,1+1,1}*[]{2,1}
\end{eqnarray}
The multiplicity in the last term is important. This factor of $2$ means that there are two possible states on the upper indices. Similarly on the lower indices, giving a total of 4 states ($2\times 2$).

Now, consider a typical Young diagram with $\ell$ boxes. Such typical states have been analyzed in \cite{Balasubramanian:2005mg}, where it was shown that there is a limit shape. 
A  typical diagram  will  be roughly symmetrical when we flip the diagram across the diagonal, and we can expect that the number of horizontal rows and vertical columns of the diagram  scale as $n_{eff}\simeq \sqrt \ell$. At this stage this is not precise. What we actually need to show is that there is a proper way of counting where $n_{eff}< c \sqrt \ell$. I will describe this counting later. For the time being, we are building an intuitive argument to get at the size of the submatrix.

In a single matrix model, one can associate different rows in the Young diagram to different eigenvalues of the matrix. This is precise when using free fermion techniques \cite{Berenstein:2004hw}. We would want to claim that there is an effective number of eigenvalues $n_{eff} \sim \sqrt \ell$ where all the physics is taking place. 
One can understand this heuristically as follows: the highest weight state for a representation of $U(N)$ with a  Young diagram with exactly $n_{eff}$ rows is invariant under $U(N-n_{eff})$, so one can claim that the corresponding $U(N-n_{eff})$ lacks any excitations.

The heuristic argument that physics resides on a submatrix needs to be made much more precise \footnote{The argument is technical and can be skipped on a first reading}.   After all, the highest weight state is not the representation.  The proper way of stating
 the information correctly  is that representations of $U(N)$ are in one to one correspondence with their highest weight states. When we consider $U(n_{eff})\times U(N-n_{eff})$ as a subgroup of $U(N)$, a highest weight state of $U(N)$ decomposes into a highest weight state of the product group. Furthermore, if we are given the Young tableaux with  $n_{eff}$ rows, the weight of the highest weight state with respect to $U(N-n_{eff})$ vanishes. This is the trivial representation for $U(N-n_{eff})$. We need to do the same argumentation with the lower indices, but we consider the ``lowest weight state" instead. This is because the lower indices are in the `antifundamental' of $U(N)$.  The highest and lowest weight state of the upper and lower indices respectively will be paired up by 
requiring invariance of the  state under the Cartan of the diagonal embedding  $U(N)\subset U(N)\times U(N)$.
Now, to get to the full gauge invariant state of the diagonal $U(N)$ and not the gauge-variant highest weight state, we need to average over the group. This is a way to implement the trace between the upper and lower indices: in the tensor product of Tableaux and it's conjugate representation there is only one trivial representation.
 This averaging can be done using coherent states. The coherent states of the representation are defined to  be group translates of the highest weight state. These are parametrized by a subset of the coset manifold $U(N)/U(N-n_{eff})$ and for each of them they have their own unbroken gauge group $U(N-n_{eff})$. We can then choose a gauge where this unbroken gauge group is in the lower $U(N-n_{eff})$ block of the matrix for each such coherent state. It is in this sense that we have localized the information of the excitations to an $n_{eff}\times n_{eff}$ sub-matrix, because  it requires a form of gauge fixing to make it work.

We still need to show that $n_{eff}<c \sqrt \ell$.  To do this, we need to consider that even though there is a limit shape for Young tableaux, it might be singular in that it can have a 
few long rows or columns that grow parametrically faster than $\sqrt \ell$.  Consider $n_1$ to be the number of rows that exceed the diagonal (the length of row $i$ is greater than or equal to $i$). Similarly consider $n_2$  to be the number of columns that exceed the diagonal (the length of column $i$ is greater than or equal to $i$).
Each of these two satisfy $n_1,n_2\leq \sqrt \ell$ as otherwise the diagram would have more boxes than $\ell$. Moreover, it is easy to convince oneself that $n_1=n_2$ (there is only one box in the diagonal with maximal distance from the origin).

We define an effective number of branes of a single tableaux as $n_{eff} = n_1$ (we could equally use $n_1+n_2$) which satisfies  $n_{eff} <c \sqrt \ell$ for all tableaux, and it scales like $\sqrt \ell$ because there is a limit shape for typical Young tableaux. This number fluctuates between different tableaux. It is reasonable to expect that in the thermodynamic limit the fluctuations of $n_{eff}$ are small relative to $n_{eff}$, so that the typical $n_{eff}$ is sufficiently well defined.

The idea now is that counting D-branes in a Young diagram should be similar to how giant gravitons and dual giant gravitons are counted. Columns count giant gravitons \cite{Balasubramanian:2001nh}, and rows count the so called dual giant gravitons \cite{Corley:2001zk}. Giant gravitons and dual giant gravitons are D-branes in the bulk  \cite{McGreevy:2000cw,Grisaru:2000zn,Hashimoto:2000zp}. The effective number of branes should minimize the counting of such objects: that is the most economical description of the physics.
Here we see that if we  divide the rows and columns into two sets, the total number of giants (long columns) and dual giants  (long rows) scales like $\sqrt \ell$ for a typical tableaux.

The total entropy for the ensemble as computed in the previous section is still given (at large $\ell$) by 
\begin{equation}
S\simeq \ell\ln 2\simeq \alpha  n_{eff}^2
\end{equation}

This scaling with $n_{eff}^2$ is suggestive. It can be thought of as a matrix model with only $n_{eff}\times n_{eff}$ matrices at a fixed temperature in a {\em deconfined phase}, where the entropy scales like the total number of degrees of freedom, while the putative commutant of these matrices $U(N-n_{eff})$ is 
actually confined.

Atypical tableaux will have far fewer rows or columns and can only give subleading contributions  in the counting of states. This follows because they would act like a $U(n_{atyp})$ field theory at fixed energy $\ell >> n_{atyp}^2$. This is like a high temperature phase for $U(n_{atyp})$ where the energy is of order $\ell \simeq n_{atyp}^2 T$ and the entropy is only $n_{atyp}^2 \log (T) $ which scales parametrically much slower than the energy.

Since in general we are in a situation where $n_{eff}<<N$, we can think of this phase as a partially deconfined phase for a submatrix set of degrees of freedom, where the fraction of eigenvalues involved in the submatrix increases with the energy. 
The analysis of counting states for more than two matrices is harder, but the idea is the same. One has to iterate the problem of taking tensor products of representations.  

The result in this section and the previous section can be interpreted in terms of the Susskind-Horowitz-Polchisnki correspondence. 
The main idea is that a sufficiently long string can smoothly become a black hole \cite{Susskind:1993w}. Or if we start with a black hole, it might dissolve into either  strings or a gas of open strings on a D-brane \cite{Horowitz:1996nw}.

The idea here is that the partially deconfined phase can be thought of as being dual, via holography, to a gas of open strings on a set of $n_{eff}$ D-branes, which is a model of a black hole. In AdS spacetime, the confinement-deconfinement transition \cite{Witten:1998zw} is dual to the Hawking-Page phase transition in gravity \cite{Hawking:1982dh}: the deconfined phase is dual to a black hole phase.
We thus argue that this partially deconfined phase should correspond to some type of black hole phase.
On the other hand, the same counting of states is obtained by a single long string.  The process of going  from the string to the black hole is smooth: there is no phase transition between the black hole phase and the single long string phase. Both descriptions not only produce the same counting of states, but they are counting the same states. The only thing that has changed is how we take into account the different configurations. 

We are in the microcanonical ensemble  below the energy that would be computed from the deconfined phase in the canonical ensemble. In standard thermodynamics this is the coexistence phase on a first order first transition.  The putative dual geometry needs to have a small black hole, not a large black hole. The large black hole has entropy and energy of order  $N^2$.  In what follows, I will describe how to utilize  this idea further  to help describe  small black holes in $AdS_5\times S^5$.

\section{The small black hole in AdS and phase separation on the sphere}\label{sec:smallbh}

Consider a thermodynamic ensemble at fixed energy for type IIB string theory in the $AdS_5\times S^5$ background, and where the energy satisfies $N^{1/4}<<E<<N^2$ (measured in global units). The lower bound is related to the  Planck scale, while requiring that the energy is much less than $N^2$ signifies that we are below the energy of the Hawking-Page phase transition. 
It has been suggested that for $E$ large enough relative to the Planck scale, the most entropic configuration is a small back hole, rather than a gas of gravitons in $AdS_5\times S^5$: it can not evaporate because the radiation in $AdS$ is confined and returns to the black hole.

A small black hole with radius much smaller than the radius of the $S^5$ should be localized. Extended black holes along the $S^5$ of small energy suffer from a Gregory-Laflamme instability. This regime should be accompanied by symmetry breaking on the sphere, from the $SO(6)$ to $SO(5)$. That is, from the gravity side we expect that the small black holes are localized 
and what we have been calling the coexistence phase should be thought of as being phase separated spatially in the geometry of the gravity dual. We would call the degrees of freedom participating in the small black hole the deconfined degrees of freedom.

As we argued in the previous sections, this should be thought of as thermal ensemble where a submatrix set of  degrees of freedom is deconfined.  This idea was first proposed in 
\cite{Asplund:2008xd}, but it is based on an incomplete description of the geometry. The purpose of this section is to get to a better understanding of why this is still a correct intuition 
in the degrees of freedom of the full quantum theory. By contrast, the analysis of Hanada and Maltz \cite{Hanada:2016pwv} shows that this can be consistent with estimates of the entropy, energy and temperature. However,  their description of the dynamics says nothing about the symmetry breaking from $SO(6)\to SO(5)$.  This needs to be fixed.

Unfortunately a direct analysis of this set of configurations is not possible at the moment. What we can do instead is realize that the small black holes can be thought of as point particles in the AdS geometry, and they can be boosted along the $S^5$ directions. That is, we can give them some $SO(6)$ R-charge by giving them a velocity. Giving a velocity to a point particle of the sphere breaks the symmetry to   $SO(2)\times SO(4)$ (if we fix the momentum, we have fixed the $SO(2)$ R-charge). We are allowed to consider such an ensemble in the 
dual field theory because the $SO(2)$ is conserved: we can fix the charge or the chemical potential for the conserved quantity. We choose to work in the fixed R-charge setup.

A small black hole that has some velocity should preserve the $SO(4)$ symmetry, but break further the $SO(2)$ by localizing somewhere on a great circle. Naively, fixing the R-charge will delocalize the state, but this is just the quantization of a collective mode of the black hole (the center of mass motion). What we need to show is that such localization phenomenon is predicted by the quantum field theory at strong coupling. First we need to show that the set of states is sufficiently compact in the geometry so that it starts looking like a point particle that sits on a geodesic trajectory in $AdS_5\times S^5$. That is, it needs to localize to a point in the great circle that is invariant under the $SO(4)$ symmetry of the $S^5$.

Secondly, we need to show that the correct set of states are localized on this great circle once we take the collective center of mass motion out of the problem.
This localization transition in the $SO(2)$ for the moving  black hole is a signal that the static black hole breaks the symmetry from $SO(6)\to SO(5)$.

On the gravity side, the small black holes (without angular momentum) have been analyzed numerically in \cite{Dias:2016eto} and have been argued to exist (numerical solutions have been found). They also match the Schwarzschild limit.  
More recently, such configurations have been analyzed dynamically in a $1/d$ expansion in \cite{Herzog:2017qwp}. 

Now, we need to argue for localization on the field theory side. There are two regimes where we can do this. The first one is the long string regime, and the second one will be given by a D-brane description of the states. We will try to argue that these give similar results. The rest of the argument 
is about estimating how the physics behaves when we have a string state with energy $E$ and $SO(2)$  R-charge $J$, in a limit where $(E-J)/ J$ is finite and small.

Let us start with the single string states at fixed $R$ charge $J$, and let us fix $E-J$ as we scale $J$. In this limit the R-charge dominates, and if we set $E-J=0$, we get a half BPS state. This state should be a single trace is given by $\Tr(Z^J)$. As we increase $E-J$, we have to add defects to the 
string. This produces an effective spin chain dynamics where defects (magnons) hop on a lattice with $J$ sites \cite{Berenstein:2002jq}. The dynamics of the spin chain is integrable \cite{Minahan:2002ve}, see \cite{Beisert:2010jr} for a full review. 
As we increase $E-J$, the number of defects grows. We are interested in the limit where $(E-J)/J$ is finite and small: this should be thought of as a rather boosted configuration of a string. We will also allow $g_{YM}^2 N$ to vary, where we will eventually go to strong coupling.

The magnon dispersion relation at momentum $p= 2\pi n/J$ in the volume is given by
\begin{equation}
(E-J)_{mag} = \sqrt{1+ \tilde \lambda\sin^2(\pi n/J)}= \sqrt{1 +  \tilde \lambda  \sin^2(p/2)}
\end{equation}
where $\tilde \lambda$ is proportional to the t' Hooft coupling $g_{YM}^2 N$.
This dispersion relation is a consequence of integrability: the single magnons are in short representations of a centrally extended algebra on the spin chain \cite{Beisert:2006qh}. 
Now, to make a state with large $E-J$, we should think of it as a gas of magnons on the lattice. When we scale $J$, we also scale  $E-J$ proportionally, so as we increase the lattice volume of the system, we are also increasing the energy of the excitations proportional to the volume. 
This is a standard thermodynamic limit at finite density. The typical state should therefore be thought of as a gas of magnons at a fixed temperature $T_{lat}$, and we expect that the entropy is also proportional to the volume. This shows that these single string states have the Hagedorn 
behavior: the entropy is proportional to the energy. Indeed, the Hagedorn temperature has been computed via integrability in \cite{Harmark:2017yrv,Harmark:2018red}

We want to analyze these configurations further. For that we need to think about varying $\tilde \lambda$, from small values to large values. In this limit, at fixed $p$, we can drop the $1$ in the dispersion relation and we find that $(E-J)_{mag}\simeq \sqrt{ \tilde \lambda}|\sin p/2 |$. 
For small $p$, this is the dispersion of a massless particle in $1+1$ dimensions. If we fix the energy density, raising $\tilde \lambda$ is equivalent to lowering the lattice temperature, and the typical $p$ decreases. It is therefore self-consistent to take the typical $p$ small. 
This can conflict with removing the $1$ from the dispersion relation. 

Having the lattice of $Z$ gives us a theory of excitations on the lattice with $SO(4)\times SO(4) \times SO(2)$ bosonic symmetry, which is the symmetry of  $SO(4,2)\times SO(6)$ that survives in the ground state. The $SO(2)$ symmetry is associated to time translation.  
 The theory is well approximated by a free theory (this can be either the plane wave limit \cite{Berenstein:2002jq}, or just a dilute approximation).
In such a theory there is no breaking of the $SO(4)\times SO(4)$ when we go to a thermal state, due to equipartition. 

Let us consider the bosons in the theory, which are $SO(4)\times SO(4)$ lattice scalars in the $4+4$ representation and call them $\phi_{1\dots 8}$. A naive estimate for the transverse size of the string is 
\begin{equation}
r_\perp^2=\vev{\sum \phi^2}
\end{equation} 
at any point in the lattice. This stays fixed as we scale $E,J$ holding $\tilde \lambda$ fixed (it is a local observable and should have a thermodynamic limit). 

Now, we need to understand the longitudinal spreading of the string along the $SO(2)$ direction. The most important realization is that $p$ is a geometric angle of the large circle. This was originally understood in a model of the strong dynamics \cite{Berenstein:2005jq}, and then 
the string solutions that correspond to this physics, called giant magnons, where constructed by Hoffman and Maldacena \cite{Hofman:2006xt}. Because the $p$ that we need are small, the angle covered by a single (giant)  magnon is small at large $\tilde \lambda$.
We can therefore interpret the longitudinal size of the string as a random walk on the circle with $k$ steps (one per magnon), each of typical size $p_{typ}$. The size of this random walk only grows as $\sqrt{k}$ and is suppressed as we increase $\tilde \lambda$: there are both fewer magnons, and
each such magnon is pushed towards smaller values of $p$ because of energy considerations.
At very small  $\tilde \lambda$ there is no penalty for having large $p$, and the random walk steps can be arbitrarily big on the circle. Such a configuration has no center of mass. On the other hand, at large $\tilde \lambda$, the string size can be made small relative to the size of the circle.
Moreover, as we increase $J$, the longitudinal size only scales as $\sqrt J$, while the transverse size is fixed. That means that the ten dimensional density of the configuration is increasing. It is reasonable to expect that this configuration on the $AdS_5\times S^5$ will transition  into a localized black hole on the circle: the string is localized and the matted density at its center of mass is increasing, as per the argument  \cite{Susskind:1993w,Horowitz:1996nw}. Notice that this localization is induced by the changing dynamics from weak to strong coupling: one can argue that at large $J$, the configuration starts to localize at a critical $\tilde \lambda$, where we go from unbroken $SO(2)$ (where the string is randomly distributed on the circle) to a localized phase. The critical transition is when the longitudinal  size of the localized string at strong coupling is comparable to the circumference of the circle itself, so we are allowed wrapping configurations in the random walk.

\begin{figure}[ht]
\begin{center}
\includegraphics[width=8cm]{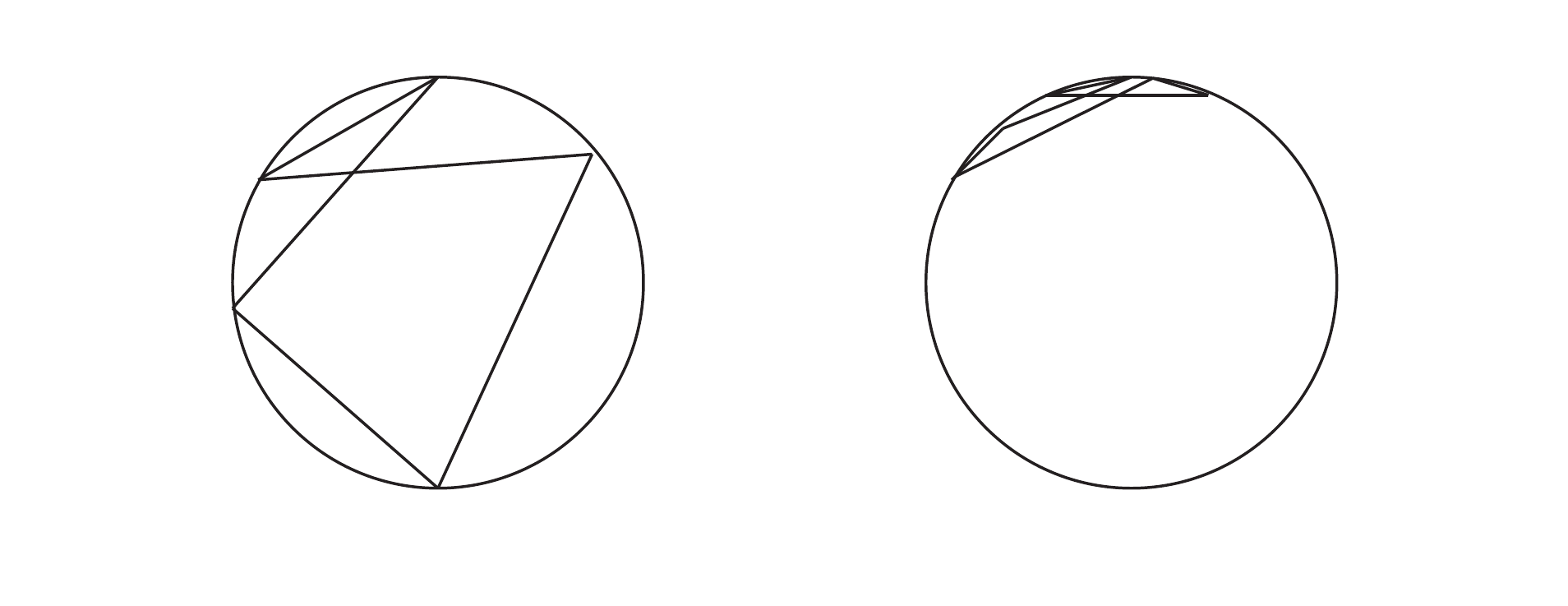}\caption{One the left, at weak coupling there is no cost in energy to have long magnons. One the right, at strong coupling the magnons are short and the string localizes.}\label{fig:weak-strong}
\end{center}
\end{figure}

A simple way to express this information is that there is an energy cost to extend the string into the $SO(2)$ direction.This is pictorially depicted in figure \ref{fig:weak-strong}

Now, to tie this long string counting of states with the previous sections we want to also give a D-brane (submatrix) picture of the relevant configurations that correspond to counting black hole states. This proceeds as follows. Since the state has a large R-charge relative to the energy it should be mostly a half BPS state. Half BPS states are completely classified in the field theory at weak coupling \cite{Corley:2001zk}. These states are protected on going from weak to strong coupling (they can not pair up with other representations of the super conformal group to make long representations). 
The basis states are charaterized by a Young tableaux and can be interpreted in terms of giant gravitons and dual giant gravitons. Geometrically, these are better understood on a 2-d phase space \cite{Berenstein:2004kk}. The ground state is a disk of fermions. Giant gravitons are holes in the Fermi sea and dual giant gravitons are single fermion excitations away from the fermi sea.  The giants and dual giants are assigned to  points in this space, where the R-charge carried by the state is related to the radial direction magnitude of the position of the object. The closer the point is to the edge of the droplet, the less R-charge it has. 
To give these giants a position that is localized, one needs a collective coordinate description of these states in the dual quantum theory. This was provided via appropriate coherent states in \cite{Berenstein:2013md}, and the states are described by evaluating generating series of Young tableaux states at particular parameters. This is similar in spirit to the theory of coherent states in a harmonic oscillator. One can consider the state $\exp(\xi a^\dagger)\ket 0$ in two different ways. For a fixed $\xi$ it describes a coherent state. For $\xi$ a formal parameter it is a generating series of all the states in the harmonic oscillator: each state $\ket M$ is captured by the coefficient of $\xi^M$ in the expansion. Multiple branes can also be analyzed (see \cite{Berenstein:2017abm} for describing these states in a particular limit).

Given two such branes, one can compute the spectrum of strings stretching between them directly from field theory. The combinatorial computation was done in \cite{deMelloKoch:2007nbd, Bekker:2007ea}. With the collective coordinates, this was done up to three loop orders in \cite{Berenstein:2013eya,Dzienkowski:2015zba}. The  spectrum of ground states coincides with the string sigma model computation \cite{Berenstein:2014isa}. This matches the states to pieces of the giant magnon \cite{Hofman:2006xt} cut at the position of the D-branes. These states are protected by supersymmetry 
by the central charge extension of the spin chain (with boundary conditions) \cite{Berenstein:2014zxa}. These states correspond to an ${\cal N}=4 $ SYM set of excitations on the worldvolume of the giant graviton. For dual giants, the computation is done in the Higgs branch, as can be understood from  \cite{Berenstein:2007zf}.

The states have a dispersion relation given by
\begin{equation}
(E-J)_{open\ string}= \sqrt{n^2 +\frac{ \tilde \lambda}4  |\xi- \xi'|^2}
\end{equation}
where $n$ is the momentum of the excitation along the volume of the brane ($n$ is along the $SO(6)$ directions for giant gravitons, and along the $SO(4,2)$ directions for dual giants).
The point is that at large separation, the open string states are massive.

To do thermodynamics, we now need to consider various scenarios. Assume we have a single giant graviton (or dual giant). We need to distribute $E-J$ on the open strings that are available. For a single giant these have $\xi=\xi'$. The counting of states on the worldvolume theory on the giant is like ${\cal N}=4 $ SYM on an $S^3$ of radius $1$, for group $U(1)$. At high energy, it is like being at high temperature, and we can use thermodynamics to assign an effective temperature $T_1$ where
\begin{equation}
E-J= \alpha T_1^4
\end{equation}
and the corresponding entropy scales as 
\begin{equation}
S\simeq T_1^3.
\end{equation}

 Assume now that we have two such giants that are separated from each other. Then
\begin{equation}
(E-J)=  2\alpha T_{2, sep}^4
\end{equation}
so that $T_{2, sep} = 2^{-1/4} T_1$
and the entropy is 
\begin{equation}
S_{2, sep} \propto  2 T_{2, sep}^3 = 2^{1/4} S_1 > S_1
\end{equation}
We can do this because the open string stretching between them are very massive: they are suppressed by the energy cost of separating the branes.
This entropy is much larger than the entropy of the first configuration.

 If the branes are on top of each other, then we need to count also the open strings between them, so that $(E-J)=  2^2\alpha T_{2, near}^4 $ and the result is instead 
\begin{equation}
S_{2, near} \propto  4 T_{2, near}^3 = (4)^{1/4} S_1 > S_{2, sep}>  S_1
\end{equation}
so that the branes have more entropy if they are on top of each other, even if the effective temperature of the excitations has gone down. For $M$ branes on top of each other, we get that 
\begin{equation}
S_{M, near} \propto M^2 T_{M, near}^3 = M^{2/4} S_1
\end{equation}
so that the system prefers to have as many branes on top of each other as is allowed.

This also precludes many additional closed strings being in the bulk: their entropy goes down as the effective temperature goes down and as $M$ increases. The dominant effect is that $M$ grows.

Each of these will carry $R$-charge given by  $J/M$, that becomes small, so the branes preferably are near the edge of the droplet. At the same time the effective temperature goes down so that one can argue that only the effective massless excitations on the branes contribute, but not massive strings stretching between these branes. Again, because the `branes' want to be on top of each other for entropic reasons, they localize in the R-charge direction: their relative coordinates on the phase space plane go to zero. 
If we make $M$ too large, then $J/M$ becomes too small and the branes spread longitudinally along the circle due to the uncertainty principle (they start delocalizing). On top of this effect, the branes act as  fermions (they behave like eigenvalues in the matrix quantum mechanics), so they can not be to close on top of each other. This can be clearly seen also in the supergravity dual geometries to these states when many of these are on top of each other. These are geometries constructed by Lin, Lunin and Maldacena  \cite{Lin:2004nb}, also known as LLM geometries.
With the extra spreading the distance along the circle between two of the branes becomes large, and the open string stretching between them become too massive so that the counting of states starts to decrease again. There should be an optimal equilibrium value of $M=n_{eff}$ that corresponds to the maximum entropy configuration. These states are localized and the reason that makes them compact on the circle is the same as for the strings: the giant magnons or the open  strings are getting heavy. These have the same dispersion relation. Notice that both are controlled by the central charge extension of the spin chain. The difference is that one is parametrized by a momentum and the other by a relative angle, but these are geometrically related to each other in the description of giant magnons \cite{Hofman:2006xt}. We really seem to be counting the same type of objects on the string and D-brane side. 

Notice that what has really been  constructed is a theory for $S_{M, near}$ for $M$ not too big: it is not a faithful description of the physics when $M$ is close to saturation. Thus the details of the putative (dual) black hole are not yet fully under control. The evidence is there that the naive phase diagram
 from gravity can be realized by dynamics in the field theory dual and that this phase is partial deconfinement characterized by an ${\cal N}=4 $ SYM thermal ensemble on a $SU(n_{eff}=M)$ field theory dynamics with $n_{eff}<<N$ and with spontaneous symmetry breaking of the $SO(2)$ symmetry.

This suggests that the picture of these states is qualitatively identical to what we argued in the previous sections: there is a counting of states by strings and by D-branes. Both countings give the same answer in the field theory variables. The counting looks very different in the gravity picture.
Notice that this result again supports the Susskind-Horowitz-Pochinski idea: we go from strings to black holes (as counted by a gas of strings on D-branes)  smoothly. In all of these computations  the number $n_{eff}$ has statistical fluctuations. The heuristic picture developed here is not advanced enough yet to 
compute these fluctuations. Again, if $n_{eff}$ is large enough we can expect that the fluctuations are small relative to $n_{eff}$ and the average value makes sense. 
Also, unlike the toy model of the previous section, here one expects that the strength of the string interactions should matter to describe the long string phase. This problem is then dominated by the self-gravity of the string and the physics of such a process has been discussed in the work \cite{Horowitz:1997jc}.

\section{Conclusion}\label{sec:conc}

In this paper I have provided strong evidence that a partial deconfinement phase exists for matrix quantum mechanical, models and in the AdS/CFT correspondence. This phase describes a coexistence phase when the confinement/deconfinement transition is of first order and it is characterized by a  submatrix set of degrees of freedom being deconfined, while the commutant of these degrees of freedom sits in the ground state. The statement is rigorous for free gauged matrix quantum mechanics.

The states can be counted either by counting long strings or by a combinatorial method using Young diagrams that has an interpretation of counting states that lie on a submatrix. The ensemble that is studied is at fixed energy, which is more appropriate for a coexistence phase.

This insight was used to provide additional evidence that the ${\cal N}=4$ SYM theory has a phase that is dual to a small black hole phase in the bulk. I showed by an entropy argument that the field theory predicts symmetry breaking of the R charge. This required using a more general microcanonical ensemble that corresponds in the bulk to a boosted black hole. The symmetry breaking is shown for these configurations. This is evidence for the $SO(6)\to SO(5)$ symmetry breaking when the R-charge vanishes. Unfortunately the techniques in this paper do not show how to deal with this scenario yet. Barring a phase 
transition when the $SO(2)$ R-charge that the state carried is reduced, at vanishing R-charge the system should still be localized. 

One of the salient features is that these systems seem to provide a direct realization of the Susskind-Horowitz-Polchinski correspondence that when strings get sufficiently excited and compact, they transition over to a black hole smoothly.

I would like to emphasize that the results of this paper show that even a simple free gauge matrix quantum mechanics for more than one matrix is a very rich system in the microcanonical ensemble. This makes it imperative to study such systems further. In particular, it would be very nice to understand better the 
combinatorial details of counting of states in Young-diagram formalism for different tableaux. 

Also, the model describing the small black hole phase in the dual SYM needs to be improved further so that rather than arguing that there is an optimal value for the number of D-branes, that one could actually calculate it. This requires further understanding of the ``effective fermion" dynamics 
of the eigenvalues that are deconfined. 

I can also speculate that some of the more extended thermal branes with $R-$ charge that are analyzed in this paper are metastable, meaning that varying the number of branes is dynamically hard to do. The reason is that to go from $M\to M+1$ branes, one needs to nucleate an extra brane of small R-charge.
this intuition is natural in gravity: they correspond to other black holes that are not the dominant entropy configuration. The extra brane that is nucleated on going from $M\to M+1$  are located near the edge of the droplet in the phase space portrait of the half -BPS states (these are on the edge of the droplet of the disk LLM geometry).  These are far away from the configuration that we are trying to grow. Before they become full branes, they remove energy from the $M$ branes and lower their entropy: the extra state is a heavy string first. The step of growing the extra brane to be large enough to contribute seems to require a rare large fluctuation.  This should make these other configurations that do not have maximal entropy  on their own metastable. The states would be dual to an extended black brane wrapping an $S^3$ of the $S^5$ geometry. The branes R-charge together with the background flux on the sphere stabilizes them from collapse in the same way that giant gravitons are stabilized from collapse.  Their horizon topology would be a spatial $S^3\times S^5$ rather than an $S^8$.

\acknowledgments

The author would like to thank D. Gross, G. Horowitz, D. Marolf, D. O'Connor  for discussions and L. Yaffe for some correspondence.
Work  supported in part by the department of Energy under grant {DE-SC} 0011702.


\begin{thebibliography}{99}

\bibitem{Gross:1980he} 
  D.~J.~Gross and E.~Witten,
  ``Possible Third Order Phase Transition in the Large N Lattice Gauge Theory,''
  Phys.\ Rev.\ D {\bf 21}, 446 (1980).
  doi:10.1103/PhysRevD.21.446


\bibitem{Hagedorn:1968zz} 
  R.~Hagedorn,
  ``Hadronic matter near the boiling point,''
  Nuovo Cim.\ A {\bf 56}, 1027 (1968).
  doi:10.1007/BF02751614
  
\bibitem{Aharony:2003sx} 
  O.~Aharony, J.~Marsano, S.~Minwalla, K.~Papadodimas and M.~Van Raamsdonk,
  ``The Hagedorn - deconfinement phase transition in weakly coupled large N gauge theories,''
  Adv.\ Theor.\ Math.\ Phys.\  {\bf 8}, 603 (2004)
  doi:10.4310/ATMP.2004.v8.n4.a1
  [hep-th/0310285].

\bibitem{Witten:1998zw} 
  E.~Witten,
  ``Anti-de Sitter space, thermal phase transition, and confinement in gauge theories,''
  Adv.\ Theor.\ Math.\ Phys.\  {\bf 2}, 505 (1998)
  doi:10.4310/ATMP.1998.v2.n3.a3
  [hep-th/9803131].


\bibitem{Hawking:1982dh} 
  S.~W.~Hawking and D.~N.~Page,
  ``Thermodynamics of Black Holes in anti-De Sitter Space,''
  Commun.\ Math.\ Phys.\  {\bf 87}, 577 (1983).
  doi:10.1007/BF01208266


\bibitem{Asplund:2008xd} 
  C.~T.~Asplund and D.~Berenstein,
  ``Small AdS black holes from SYM,''
  Phys.\ Lett.\ B {\bf 673}, 264 (2009)
  doi:10.1016/j.physletb.2009.02.043
  [arXiv:0809.0712 [hep-th]].


\bibitem{Hanada:2016pwv} 
  M.~Hanada and J.~Maltz,
  ``A proposal of the gauge theory description of the small Schwarzschild black hole in AdS$_5\times$S$^5$,''
  JHEP {\bf 1702}, 012 (2017)
  doi:10.1007/JHEP02(2017)012
  [arXiv:1608.03276 [hep-th]].

\bibitem{Yaffe:2017axl} 
  L.~G.~Yaffe,
  ``Large $N$ phase transitions and the fate of small Schwarzschild-AdS black holes,''
  Phys.\ Rev.\ D {\bf 97}, no. 2, 026010 (2018)
  doi:10.1103/PhysRevD.97.026010
  [arXiv:1710.06455 [hep-th]].

\bibitem{Berenstein:2005aa} 
  D.~Berenstein,
  ``Large N BPS states and emergent quantum gravity,''
  JHEP {\bf 0601}, 125 (2006)
  doi:10.1088/1126-6708/2006/01/125
  [hep-th/0507203].

\bibitem{Berenstein:2015ooa} 
  D.~Berenstein,
  ``Extremal chiral ring states in the AdS/CFT correspondence are described by free fermions for a generalized oscillator algebra,''
  Phys.\ Rev.\ D {\bf 92}, no. 4, 046006 (2015)
  doi:10.1103/PhysRevD.92.046006
  [arXiv:1504.05389 [hep-th]].

\bibitem{Balasubramanian:2004nb} 
  V.~Balasubramanian, D.~Berenstein, B.~Feng and M.~x.~Huang,
  ``D-branes in Yang-Mills theory and emergent gauge symmetry,''
  JHEP {\bf 0503}, 006 (2005)
  doi:10.1088/1126-6708/2005/03/006
  [hep-th/0411205].

\bibitem{deMelloKoch:2007rqf} 
  R.~de Mello Koch, J.~Smolic and M.~Smolic,
  ``Giant Gravitons - with Strings Attached (I),''
  JHEP {\bf 0706}, 074 (2007)
  doi:10.1088/1126-6708/2007/06/074
  [hep-th/0701066].



\bibitem{Berenstein:2004hw} 
  D.~Berenstein,
  ``A Matrix model for a quantum Hall droplet with manifest particle-hole symmetry,''
  Phys.\ Rev.\ D {\bf 71}, 085001 (2005)
  doi:10.1103/PhysRevD.71.085001
  [hep-th/0409115].

\bibitem{Bhattacharyya:2008rb} 
  R.~Bhattacharyya, S.~Collins and R.~de Mello Koch,
  ``Exact Multi-Matrix Correlators,''
  JHEP {\bf 0803}, 044 (2008)
  doi:10.1088/1126-6708/2008/03/044
  [arXiv:0801.2061 [hep-th]].

\bibitem{Mattioli:2016eyp} 
  P.~Mattioli and S.~Ramgoolam,
  ``Permutation Centralizer Algebras and Multi-Matrix Invariants,''
  Phys.\ Rev.\ D {\bf 93}, no. 6, 065040 (2016)
  doi:10.1103/PhysRevD.93.065040
  [arXiv:1601.06086 [hep-th]].


\bibitem{Ramgoolam:2016ciq} 
  S.~Ramgoolam,
  ``Permutations and the combinatorics of gauge invariants for general N,''
  PoS CORFU {\bf 2015}, 107 (2016)
  doi:10.22323/1.263.0107
  [arXiv:1605.00843 [hep-th]].


\bibitem{Pasukonis:2013ts} 
  J.~Pasukonis and S.~Ramgoolam,
  ``Quivers as Calculators: Counting, Correlators and Riemann Surfaces,''
  JHEP {\bf 1304}, 094 (2013)
  doi:10.1007/JHEP04(2013)094
  [arXiv:1301.1980 [hep-th]].


\bibitem{Balasubramanian:2001nh} 
  V.~Balasubramanian, M.~Berkooz, A.~Naqvi and M.~J.~Strassler,
  ``Giant gravitons in conformal field theory,''
  JHEP {\bf 0204}, 034 (2002)
  doi:10.1088/1126-6708/2002/04/034
  [hep-th/0107119].

\bibitem{Corley:2001zk} 
  S.~Corley, A.~Jevicki and S.~Ramgoolam,
  ``Exact correlators of giant gravitons from dual N=4 SYM theory,''
  Adv.\ Theor.\ Math.\ Phys.\  {\bf 5}, 809 (2002)
  doi:10.4310/ATMP.2001.v5.n4.a6
  [hep-th/0111222].


\bibitem{McGreevy:2000cw} 
  J.~McGreevy, L.~Susskind and N.~Toumbas,
  ``Invasion of the giant gravitons from Anti-de Sitter space,''
  JHEP {\bf 0006}, 008 (2000)
  doi:10.1088/1126-6708/2000/06/008
  [hep-th/0003075].


\bibitem{Grisaru:2000zn} 
  M.~T.~Grisaru, R.~C.~Myers and O.~Tafjord,
  ``SUSY and goliath,''
  JHEP {\bf 0008}, 040 (2000)
  doi:10.1088/1126-6708/2000/08/040
  [hep-th/0008015].


\bibitem{Hashimoto:2000zp} 
  A.~Hashimoto, S.~Hirano and N.~Itzhaki,
  ``Large branes in AdS and their field theory dual,''
  JHEP {\bf 0008}, 051 (2000)
  doi:10.1088/1126-6708/2000/08/051
  [hep-th/0008016].

\bibitem{Balasubramanian:2005mg} 
  V.~Balasubramanian, J.~de Boer, V.~Jejjala and J.~Simon,
  ``The Library of Babel: On the origin of gravitational thermodynamics,''
  JHEP {\bf 0512}, 006 (2005)
  doi:10.1088/1126-6708/2005/12/006
  [hep-th/0508023].

\bibitem{Susskind:1993ws} 
  L.~Susskind,
  ``Some speculations about black hole entropy in string theory,''
  In *Teitelboim, C. (ed.): The black hole* 118-131
  [hep-th/9309145].


\bibitem{Horowitz:1996nw} 
  G.~T.~Horowitz and J.~Polchinski,
  ``A Correspondence principle for black holes and strings,''
  Phys.\ Rev.\ D {\bf 55}, 6189 (1997)
  doi:10.1103/PhysRevD.55.6189
  [hep-th/9612146].

\bibitem{Dias:2016eto} 
  \v O.~J.~C.~Dias, J.~E.~Santos and B.~Way,
  ``Localised $\bf{AdS_5\times S^5}$ Black Holes,''
  Phys.\ Rev.\ Lett.\  {\bf 117}, no. 15, 151101 (2016)
  doi:10.1103/PhysRevLett.117.151101
  [arXiv:1605.04911 [hep-th]].

\bibitem{Herzog:2017qwp} 
  C.~P.~Herzog and Y.~Kim,
  ``The Large Dimension Limit of a Small Black Hole Instability in Anti-de Sitter Space,''
  JHEP {\bf 1802}, 167 (2018)
  doi:10.1007/JHEP02(2018)167
  [arXiv:1711.04865 [hep-th]].

\bibitem{Berenstein:2002jq} 
  D.~E.~Berenstein, J.~M.~Maldacena and H.~S.~Nastase,
  ``Strings in flat space and pp waves from N=4 superYang-Mills,''
  JHEP {\bf 0204}, 013 (2002)
  doi:10.1088/1126-6708/2002/04/013
  [hep-th/0202021].

\bibitem{Minahan:2002ve} 
  J.~A.~Minahan and K.~Zarembo,
  ``The Bethe ansatz for N=4 superYang-Mills,''
  JHEP {\bf 0303}, 013 (2003)
  doi:10.1088/1126-6708/2003/03/013
  [hep-th/0212208].


\bibitem{Beisert:2010jr} 
  N.~Beisert {\it et al.},
  ``Review of AdS/CFT Integrability: An Overview,''
  Lett.\ Math.\ Phys.\  {\bf 99}, 3 (2012)
  doi:10.1007/s11005-011-0529-2
  [arXiv:1012.3982 [hep-th]].

\bibitem{Beisert:2006qh} 
  N.~Beisert,
  ``The Analytic Bethe Ansatz for a Chain with Centrally Extended su(2|2) Symmetry,''
  J.\ Stat.\ Mech.\  {\bf 0701}, P01017 (2007)
  doi:10.1088/1742-5468/2007/01/P01017
  [nlin/0610017 [nlin.SI]].


\bibitem{Harmark:2017yrv} 
  T.~Harmark and M.~Wilhelm,
  ``Hagedorn Temperature of AdS$_5$/CFT$_4$ via Integrability,''
  Phys.\ Rev.\ Lett.\  {\bf 120}, no. 7, 071605 (2018)
  doi:10.1103/PhysRevLett.120.071605
  [arXiv:1706.03074 [hep-th]].

\bibitem{Harmark:2018red} 
  T.~Harmark and M.~Wilhelm,
  ``The Hagedorn temperature of AdS$_5$/CFT$_4$ at finite coupling via the Quantum Spectral Curve,''
  arXiv:1803.04416 [hep-th].


\bibitem{Berenstein:2005jq} 
  D.~Berenstein, D.~H.~Correa and S.~E.~Vazquez,
  ``All loop BMN state energies from matrices,''
  JHEP {\bf 0602}, 048 (2006)
  doi:10.1088/1126-6708/2006/02/048
  [hep-th/0509015].

\bibitem{Hofman:2006xt} 
  D.~M.~Hofman and J.~M.~Maldacena,
  ``Giant Magnons,''
  J.\ Phys.\ A {\bf 39}, 13095 (2006)
  doi:10.1088/0305-4470/39/41/S17
  [hep-th/0604135].

\bibitem{Berenstein:2004kk} 
  D.~Berenstein,
  ``A Toy model for the AdS / CFT correspondence,''
  JHEP {\bf 0407}, 018 (2004)
  doi:10.1088/1126-6708/2004/07/018
  [hep-th/0403110].
  
\bibitem{Berenstein:2013md} 
  D.~Berenstein,
  ``Giant gravitons: a collective coordinate approach,''
  Phys.\ Rev.\ D {\bf 87}, no. 12, 126009 (2013)
  doi:10.1103/PhysRevD.87.126009
  [arXiv:1301.3519 [hep-th]].
  
\bibitem{Berenstein:2017abm} 
  D.~Berenstein and A.~Miller,
  ``Superposition induced topology changes in quantum gravity,''
  JHEP {\bf 1711}, 121 (2017)
  doi:10.1007/JHEP11(2017)121
  [arXiv:1702.03011 [hep-th]].
  
\bibitem{deMelloKoch:2007nbd} 
  R.~de Mello Koch, J.~Smolic and M.~Smolic,
  ``Giant Gravitons - with Strings Attached (II),''
  JHEP {\bf 0709}, 049 (2007)
  doi:10.1088/1126-6708/2007/09/049
  [hep-th/0701067].
  
  
\bibitem{Bekker:2007ea} 
  D.~Bekker, R.~de Mello Koch and M.~Stephanou,
  ``Giant Gravitons - with Strings Attached. III.,''
  JHEP {\bf 0802}, 029 (2008)
  doi:10.1088/1126-6708/2008/02/029
  [arXiv:0710.5372 [hep-th]].
  
\bibitem{Berenstein:2013eya} 
  D.~Berenstein and E.~Dzienkowski,
  ``Open spin chains for giant gravitons and relativity,''
  JHEP {\bf 1308}, 047 (2013)
  doi:10.1007/JHEP08(2013)047
  [arXiv:1305.2394 [hep-th]].
  
\bibitem{Dzienkowski:2015zba} 
  E.~Dzienkowski,
  ``Excited States of Open Strings From $\mathcal{N}=4$ SYM,''
  JHEP {\bf 1512}, 036 (2015)
  doi:10.1007/JHEP12(2015)036
  [arXiv:1507.01595 [hep-th]].
  
\bibitem{Berenstein:2014isa} 
  D.~Berenstein and E.~Dzienkowski,
  ``Giant gravitons and the emergence of geometric limits in beta-deformations of $ \mathcal{N}=4 $ SYM,''
  JHEP {\bf 1501}, 126 (2015)
  doi:10.1007/JHEP01(2015)126
  [arXiv:1408.3620 [hep-th]].
  
\bibitem{Berenstein:2014zxa} 
  D.~Berenstein,
  ``On the central charge extension of the $ \mathcal{N}=4 $ SYM spin chain,''
  JHEP {\bf 1505}, 129 (2015)
  doi:10.1007/JHEP05(2015)129
  [arXiv:1411.5921 [hep-th]].
  
\bibitem{Berenstein:2007zf} 
  D.~Berenstein and S.~E.~Vazquez,
  ``Giant magnon bound states from strongly coupled N=4 SYM,''
  Phys.\ Rev.\ D {\bf 77}, 026005 (2008)
  doi:10.1103/PhysRevD.77.026005
  [arXiv:0707.4669 [hep-th]].
  
\bibitem{Lin:2004nb} 
  H.~Lin, O.~Lunin and J.~M.~Maldacena,
  ``Bubbling AdS space and 1/2 BPS geometries,''
  JHEP {\bf 0410}, 025 (2004)
  doi:10.1088/1126-6708/2004/10/025
  [hep-th/0409174].
  
\bibitem{Horowitz:1997jc} 
  G.~T.~Horowitz and J.~Polchinski,
  ``Selfgravitating fundamental strings,''
  Phys.\ Rev.\ D {\bf 57}, 2557 (1998)
  doi:10.1103/PhysRevD.57.2557
  [hep-th/9707170].
  
\end{thebibliography}
\end{document}